# Neutron imaging with a Micromegas detector

F. Jeanneau, R. Junca, J. Pancin, M. Voytchev, S Andriamonje, V. Dangendorf, I. Espagnon,
H. Friedrich, A. Giganon, I. Giomataris, A. Menelle, A. Pluquet and L. R. Rodríguez

*Abstract*— The micropattern gaseous detector Micromegas has been developed for several years in Saclay and presents good performance for neutron detection. A prototype for neutron imaging has been designed and new results obtained in thermal neutron beams are presented. Based on previous results demonstrating a good 1D spatial resolution, a tomographic image of a multiwire cable has been performed using a 1D Micromegas prototype. The number of pillars supporting the micromesh is too large and leads to local losses of efficiency that distort the tomographic reconstruction. Nevertheless, this first tomographic image achieved with this kind of detector is very encouraging.

The next worthwhile development for neutron imaging is to achieve a bi-dimensional detector, which is presented in the second part of this study. The purpose of measurements was to investigate various operational parameters to optimize the spatial resolution. Through these measurements the optimum spatial resolution has been found to be around 160 µm (standard deviation) using Micromegas operating in double amplification mode. Several 2D imaging tests have been carried out. Some of these results have revealed fabrication defects that occurred during the manufacture of Micromegas and that are limiting the full potential of the present neutron imaging system.

## I. Introduction

Neutronography is an essential technique in order to achieve the accuracy necessary for the quality controls imposed in the modern industry (for example: aerospace programs). The most used devices up to now are based on CCD cameras coupled with neutron-sensitive scintillators which offer a spatial resolution around 200 µm (FWHM) [1]. Another technique is the Imaging Plate, giving also a good spatial resolution but requiring an independent laser system for the image development [2]. In this paper we propose a device for neutron imaging based on Micromegas (MICRO-MEsh GAseous Structure), a gaseous detector that has been in development in Saclay since 1996 [3] for fundamental research, in particular for the detection and tracking of charged particles. Several studies have already been performed with this detector, demonstrating its capacity to work at high neutron flux rates with a good spatial resolution [4]-[5]. Moreover this detector presents some advantages in terms of insensitivity to gamma radiations, robustness and implementation. In this study, the performance of this detector is evaluated for applications in thermal neutron imaging.

In the first part of this article, the neutron detection principle is explained from a general point of view. Then two different studies, concerning 1D or 2D prototypes, are presented. Previous investigations using a 1D Micromegas have given promising results in terms of spatial resolution (60 µm) [6] or fabrication processes [7]. The first study is dedicated to tomographic imaging with a 1D Micromegas prototype.

In a second study, these investigations are extended to explore the 2D neutron imaging capability of the Micromegas detector and to measure the optimal spatial resolution achievable with a bi-dimensional prototype.

## II. THE MICROMEGAS DETECTOR

### A. Detector principle

When charged particles travel through the gas (typically a mixture of Argon or Helium and a small percentage of isobutane), electrons are released between the drift electrode and the micromesh via a process of ionization. An electric field (~1 kV/cm) applied in the conversion gap (several hundreds of microns to few millimeters) then causes these electrons to drift towards the micromesh. After passing through the micromesh, they undergo multiplication in the narrow amplification gap (several tens of microns, see Fig. 1) where the field strength is much higher (~100 kV/cm).

However, for neutron detection, a neutron/charged particle converter is necessary to produce ionization. The conversion reactions used are neutron absorption on $^6$Li (or $^6$LiF) or $^{10}$B (respectively with cross section of 940 or 3840 barns), deposited on the drift electrode:

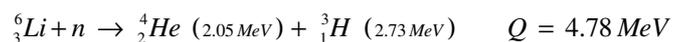

$$^6_3Li + n \rightarrow {}^4_2He\,(2.05\,MeV) + {}^3_1H\,(2.73\,MeV) \qquad Q = 4.78\,MeV$$

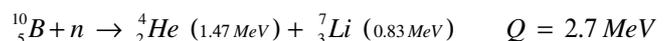

$$^{10}_5B + n \rightarrow {}^4_2He\,(1.47\,MeV) + {}^7_3Li\,(0.83\,MeV) \qquad Q = 2.7\,MeV$$

Since the incoming neutron energy is negligible compared to the reaction $Q$ value, the two reaction products (alpha, triton or

This work was supported in part by the Conseil Régional d'Ile de France.

F. Jeanneau, R. Junca, M. Voytchev, I. Espagnon, A. Pluquet are with the CEA-Saclay, DRT/DeTeCS, 91191 Gif-sur-Yvette Cedex, France. (email-address: fabien.jeanneau@cea.fr)

J. Pancin, S. Andriamonje, A. Giganon, I. Giomataris are with the CEA-Saclay, DSM/DAPNIA, 91191 Gif sur Yvette Cedex, France

V. Dangendorf and H. Friedrich are with the PTB, Braunschweig, Germany

A. Menelle is with the CEA-Saclay, DSM/DRECAM/LLB, 91191 Gif sur Yvette Cedex, France

L.R. Rodríguez is with the Dpto de Física, Universidad de Burgos, Avda. Cantabria s/n., 09006 Burgos, Spain

$^7$Li) are emitted in opposite directions [8]. The optimal thickness of the different converters (Table 1) has been estimated via a Monte-Carlo simulation using data from the ion transport code TRIM [9]. In the case of a $^6$Li converter, the total efficiency as a function of the thickness is given in Figure 2. It is a compromise between the conversion efficiency (proportional to the converter thickness) and the probability to exit from the converter material for each reaction products. The best total efficiency in the case of $^6$Li is around 20 % for a thickness of 100 µm. Since products are emitted isotropically, the conversion gap is reduced to a few hundred micrometers, in order to minimize the signal expanse over the strips for large angles.

The detector allows two modes of operation: in normal mode, amplification only takes place in the lower gap, while in pre-amplification mode, the electric field applied in the conversion gap is high enough to amplify the signal, so that the global detector gain is a contribution from the two stages.

### B. Electronics and data acquisition system

For each event, the spatial distribution of the deposited charge on the strips is collected by 4 GASSIPLEX [10] boards (96 channels per board) performing charge preamplification and multiplexing. A VME device, composed of a sequencer (*CAEN* [11] model V551) and 2 CRAMs (*CAEN* model V550 - 10 bits), is used to command the GASSIPLEX and store its signals. The acquisition system is triggered by the amplified micromesh signal. Once all the ADC values have been stored in the CRAMs, the data are transferred to a PC via a VME-MXI-2 card (National Instruments). The whole system is managed by a software based on *LabView* [12], previously developed for the CAST (Cern Axion Solar Telescope) experiment [13], allowing a pedestal suppression in order to acquire data with a higher frequency (around 2.5 kHz).

### III. TOMOGRAPHY WITH A 1D MICROMEGAS DETECTOR

### A. Setup description

Tomography measurements have been performed in March 2004 at GKSS in Geesthacht (Germany) at the GeNF (Geesthacht Neutron Facility) reactor which is essentially dedicated to neutron spectroscopy. We used the POLDI area (POLarised DIffractometer) [14] delivering a thermal neutron flux of $8.10^4$ n.cm$^{-2}$.s$^{-1}$ with an optimal beam divergence (L/D ~ 2000, with L the distance between the source and the detector and D the aperture diameter of the source). A tomographic image of a multi-wire cable (Ø 6 mm), containing 12 wires (Ø 0.5 mm each), has been achieved using a 1D Micromegas equipped with a $^6$Li converter. The readout plane contains 384 strips with a pitch of 100 µm. The conversion gap and the amplification gap (defined by the pillar's height) are 700 µm and 100 µm thick, respectively. A 1 mm wide slit is located between the object and the detector in order to reduce scattered neutrons (see Figure 3). A remotely controlled motor holds and rotates the object between each projection.

### B. Data analysis

Once the Gaussian distribution of the electronic noise is registered for each channel (obtained without neutron beam and using a random trigger), the mean value of the distribution gives the pedestal value, and the standard deviation (sd) is used to discriminate the signal from the noise. These values are registered in the acquisition system so that it only records the channels with an ADC value larger than the pedestal value plus typically 3 sd. Empty channels are then suppressed before registration, minimizing the data flow and allowing the acquisition system to run at higher frequency.

The data file contains the event number and the list of the fired strips with their ADC value. Every contiguous strip with a signal is stored in a cluster by a C++ program. Wherever strips are identified as being dead, the mean value of the neighboring nodes is inserted.

The main cluster is the one containing the strip with the maximum charge. Several parameters are extracted from this cluster: charge and number of the maximum strip, total number of strips (cluster size), total charge (sum of all ADC values) and the charge centroid, related to the incident neutron position. By selecting events with no saturation strips and low cluster size (in order to limit the error on the centroid calculation), these parameters allow the reconstruction of the incident neutron beam on the plane of strips and to improve the spatial resolution.

### C. Tomography test results

The total number of projections is 81 on 360°, with a counting rate around 200 neutrons per second and per projection. Since time allocated for each projection is about 1000 seconds, a tomographic image can be performed in about 26 hours.

Due to small data sample sizes, it was not possible to apply the event selection necessary to reach the optimal resolution (keeping only the low cluster size events) which is around 60 µm [4]. Thresholds in our case are: a multiplicity below 12 strips, and in addition events with an excessively high deposited charge per strip are rejected, in order to select the events perpendicular to the direction of the strips (25% of events are kept with this cuts set). These cuts ensure a spatial resolution of around 100 µm, calculated with the image of a slit of 500 µm wide parallel to the strips.

An image of the beam with no object, $I_0(x)$, is needed for the tomographic reconstruction [15] to monitor the beam and the defects of the detector. Each projection $I^i(x)$ is normalized with the empty run, by adjusting the number of events on the edges of the object, so that the beam or acquisition time variations are corrected. Then the ratio $\ln(I_0(x)/I^i(x))$ gives an evaluation of the integral of the neutron attenuation coefficient in the projection direction, as depicted in Figure 4 for the projection at 0°. Some structures due to the wires in the cable are visible as well as artifacts caused by the pillars sustaining the micromesh. The latter effect is also disturbing the sinogram (Figure 5 – left), corresponding to the piling of the projections, in the shape of straight lines of attenuation. The reconstruction image is represented on Figure 5 (right). Some wires and the sheath are well reconstructed but the image center is quite blurred. As a comparison, the section of the same object obtained with imaging plate (FUJIFILM BAS5000) [4] is better (Figure 5 – bottom). A spatial resolution around 50 µm was obtained with this detector.

The main reason to explain this difference of quality is certainly the high number of pillars on the plane of strips, deposited every 0.6 mm, which could be decreased by a factor of 4. Local inefficiencies due to pillars are interpreted as concentric circles, degrading the spatial resolution, especially near the rotation center. Moreover, because of the lack of statistics, the optimal set of cuts cannot be applied to achieve the best spatial resolution. However, this could be improved using a higher neutron flux.

## IV. 2D IMAGING STUDY

Tests and measurements have been carried out at the Orphée reactor at CEA-Saclay. The detector is placed in a 0.025 eV neutron beam of $10^6$ n.cm$^{-2}$.s$^{-1}$, after two collimators (B$_4$C and Cd) defining the working area.

### A. Prototype description

The prototype is the same as the 1D Micromegas except for the plane of strips. The 2D readout plane is comprised of coppered Kapton where strips are etched with a 300 µm pitch in each direction (between two circles – see Figure 6). Strips consist of hexagonal pixels connected, on the upper face for one direction and on the other face through metal-filled holes for the other direction. The 384 electronics channels are shared between the two directions (2 × 192 strips) defining a square working zone of about 6 × 6 cm². The conversion and amplification gaps are equal to 700 and 50 µm, respectively.

### B. Data analysis

The principle of data analysis is the same as for the 1D imaging prototype except for the event selection. In this case, only events with at least one cluster in X and Y, a cluster size equal to 2 or 3 strips and no saturation strips are used to evaluate the optimum spatial resolution.

### C. Estimation of the spatial resolution

In order to determine the spatial resolution of the detector, an image of a hole (∅ 0.6 mm) in a Cadmium foil (thickness 0.8 mm) is performed. This image is fitted by the convolution of a square function and a Gaussian function:

$$T(x,y) = p + A \left[ Erf\left(\frac{a+\sqrt{(x-\mu_x)^2+(y-\mu_y)^2}}{\sqrt{2}\sigma}\right) + Erf\left(\frac{a-\sqrt{(x-\mu_x)^2+(y-\mu_y)^2}}{\sqrt{2}\sigma}\right) \right]$$

where $p$ is the noise level, $A$ a constant, $a$ the half width of the hole, ($\mu_x$, $\mu_y$) the center of the hole and $\sigma$ the Gaussian sd (spatial resolution value).

### D. Results

Two different operational modes of the Micromegas have been used in these studies. Better results, in terms of spatial resolution, are obtained in pre-amplification mode. This is explained by a better reconstruction of the neutron position due to the shape of the clusters (see Figure 7). In normal mode, electrons created along the way of the particle in the conversion gap lead to a uniform signal. In pre-amplification mode, electrons created at the beginning of the conversion gap, and corresponding exactly to the neutron position, have a bigger contribution to the signal. The cluster centroid is then shifted towards the particle entry point, which improves the spatial resolution.

Figure 8 shows the image of the 0.6 mm holes. The detector plane contained some dead strips and short circuits, so the spatial resolution study was carried out using a small working zone (10×10 mm² centered on the images of the holes). The brighter spot cast by one hole was distorted because a dead strip passed through it, so the second brightest spot, coinciding with a fully operational area of the detector, was used in the following analysis.

Figure 9 shows two examples of the approximation of the hole image by the $T(x,y)$ function for different sets of event selections (essentially on the cluster size). These results have been obtained in pre-amplification. The spatial resolution can be optimized at the expense of absolute efficiency, by selecting events of lower cluster size (and therefore lower positional uncertainty). This improves the spatial resolution from 300 µm, for a 90% event acceptance rate – see Figure 9 (top), to around 160 µm if only 15% of the events are kept (cluster sizes equal to 2 or 3 in the two directions) – see Figure 9 (bottom).

In order to perform an image of a real object, a new prototype has been mounted with a new plane of strips. The object consists of a Gadolinium foil of 100 µm thick on which the letters "CEA" have been etched. Figure 10 shows the scheme of the etching and the image obtained in neutron beam. The three letters of the object are correctly reconstructed, and the dimensions are preserved, but the image quality is not very good because of inefficient areas. After careful electrical testing of the plane of strips, we found that etching defects led to some cross-talk (due to capacitive coupling) between the channels of the 2 directions, causing significant signal losses on some strips. The cluster centroid is then affected by this effect which degrades spatial resolution and response homogeneity. Nevertheless these results are very encouraging and indicate that special attention is needed during the manufacturing of the readout strips. By doing so, we will achieve the full 2D neutron imaging potential of the present detector.

V. CONCLUSION

A tomographic image has been carried out with a 1D prototype near the GKSS reactor in Geesthacht. The image of a multi-wire cable obtained after reconstruction does not present the same quality as the one obtained with the imaging plate. The main reason is the high number of pillars on the plane of strips, leading to local inefficiencies and then to artifacts in the reconstruction process. Moreover, it was not possible to apply the optimal set of cuts due to the lack of statistics. The results obtained are very encouraging despite the technical problems encountered. The number of pillars can be significantly reduced in the manufacturing process but the solution is to deposit the pillars between the strips to avoid local inefficiencies. Finally, a higher neutron flux is needed to perform new measurements and to determine the relevance of this kind of device for neutron imaging.

The 2D Micromegas prototype equipped with its acquisition electronics worked well during the experiments near Orphée reactor. A 2D spatial resolution of around 160 µm (in terms of standard deviation) has been reached in pre-amplification mode and around 180 µm in normal mode, with an acceptance level of 15 % (cluster size equal to 2 or 3 in X and in Y). Taking into account the dead strips, these results are very encouraging.

An image has been acquired with a new plane of strips. It highlighted the presence of cross-talk between the 2 directions of the plane of strips. In future developments, the manufacturing process will be optimized to avoid this kind of defect which degrades the spatial resolution and leads to local inefficiencies, manifest by some strips losing the main part of their signal. Further measurements are therefore necessary.

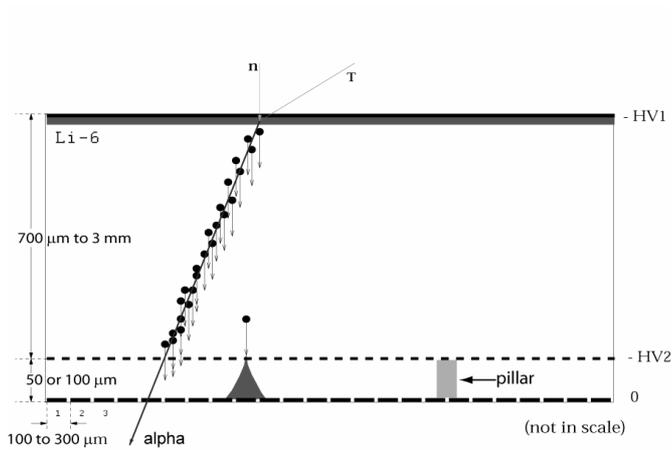

Fig. 1. Micromegas principle for neutron detection.

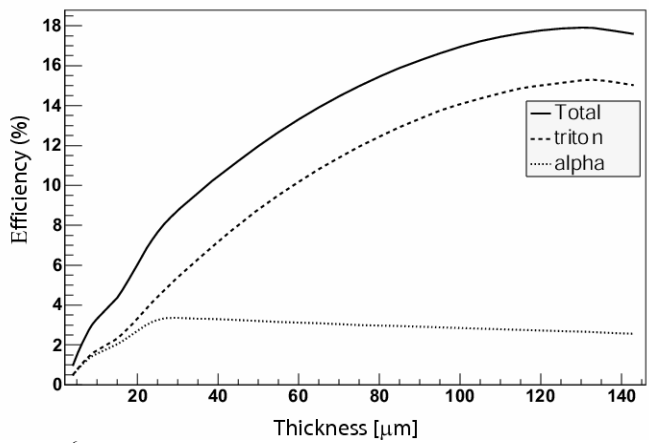

Fig. 2. $^6$Li conversion efficiency as a function of the thickness

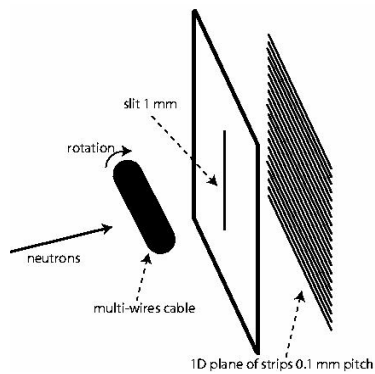

Fig. 3. Schematic view of the tomographic setup (not on scale)

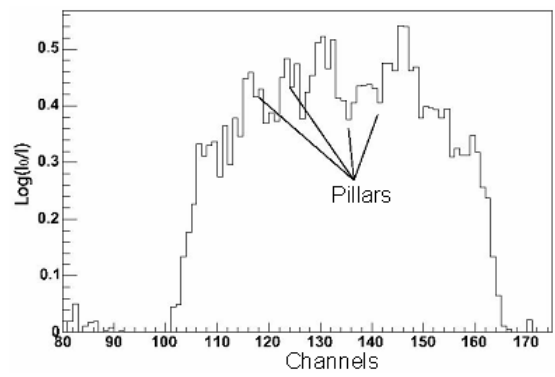

Fig. 4. Projection at 0° after normalization.

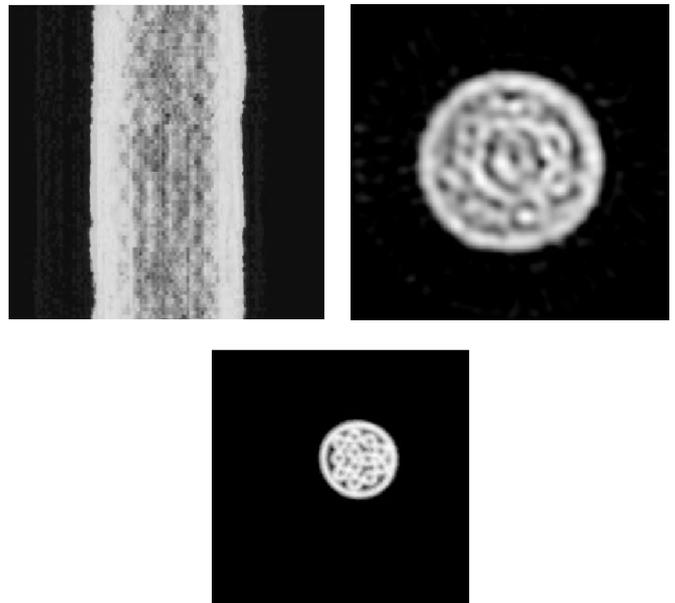

Fig. 5. Tomographic image of a multi-wires cable (∅ 6 mm, 12 wires, ∅ 0.5 mm each). Left: sinogram of the 81 projections. Right: corresponding reconstructed tomographic image using the Micromegas detector. Bottom: tomographic image using image plates.

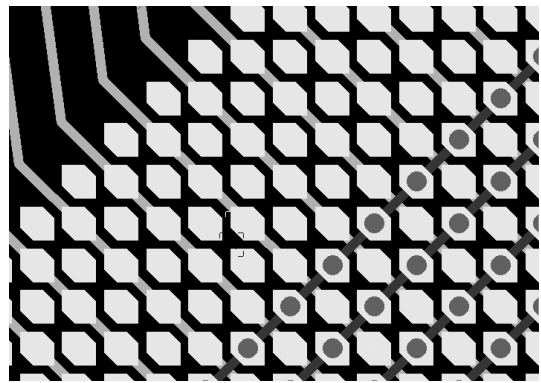

Fig. 6. Diagram of the 2D readout (pitch between two circles along the strips = 300µm)

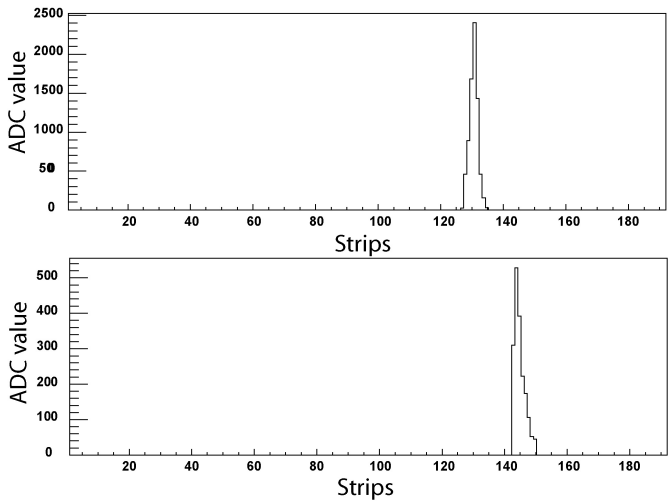

Fig. 7. Shape of the charge cluster in normal mode (top) and in pre-amplification mode (bottom)

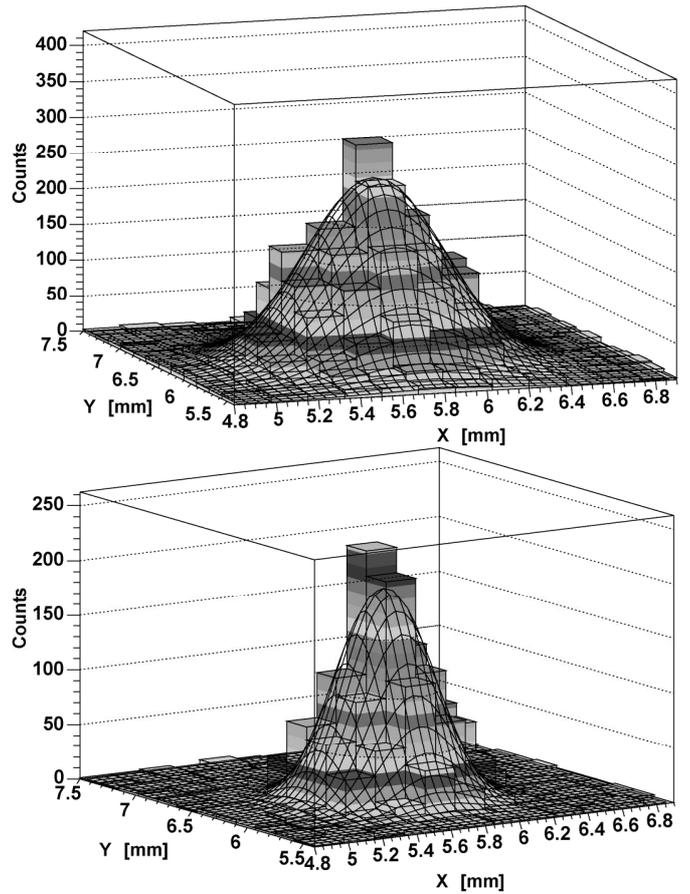

Fig. 9. Approximation of the hole with the function $T(x,y)$ for two differents set of cuts on the clusters size $n_x$ and $n_y$: (top) $n_x>1$ and $n_y>1$, (bottom) $1<n_x<4$ and $1<n_y<4$.

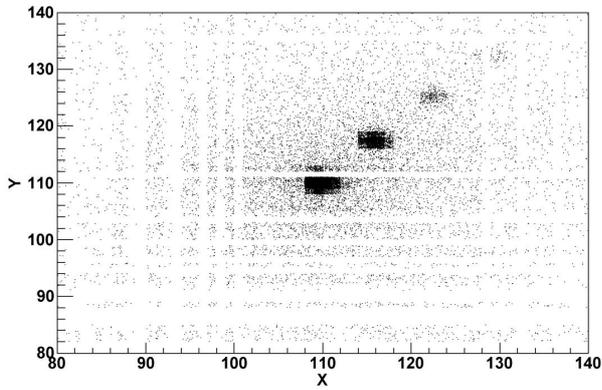

Fig. 8. Raw images of 0.6 mm holes.

TABLE 1
DIFFERENT CONVERTERS PARAMETERS

| Type | $^6$Li [1] | $^6$LiF [2] | $^{10}$B [3] |
|---|---|---|---|
| Thickness [µm] | 100 | 2 or 24 | 0.8 |
| Efficiency (%) | 18 | 1.1 or 5.9 | 3.6 |
| Optimal thickness [µm] | 130 | 33 | 3.4 |
| Max efficiency (%) | 18 | 6.4 | 7.8 |

Evaporated in:
(1) Physikalisch Technische Bundesanstalt (PTB), Braunschweig, Germany
(2) University of Burgos, Spain
(3) CERN, Geneva, Switzerland

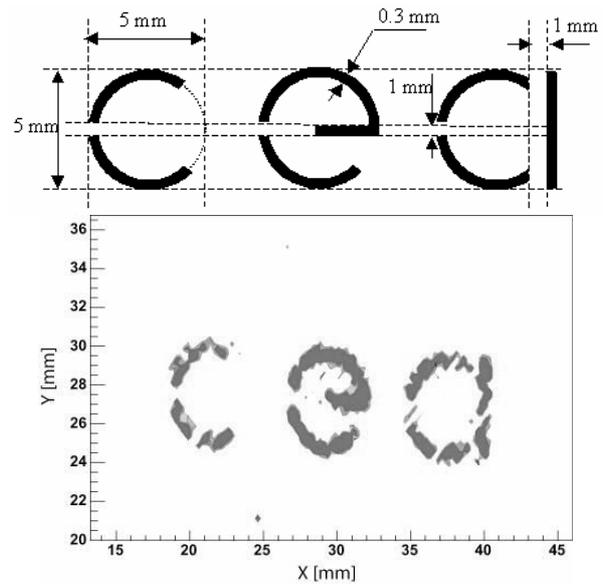

Fig. 10. Scheme of the gadolinium foil (100 µm) etching and image obtained with the detector.